\documentclass[11pt, a4paaper, oneside]{article}
\usepackage{lineno,hyperref}

\usepackage[margin=1in]{geometry}  
\usepackage{graphicx}              
\usepackage{epsfig}
\usepackage{amsmath}               
\usepackage{amssymb}
\usepackage{epstopdf}
\usepackage{verbatim}
\usepackage{mathptmx}
\usepackage{mathalpha}
\usepackage{mathtools}
\usepackage{graphicx, times}
\usepackage{amsfonts}              
\usepackage{amsthm}                
\usepackage{multicol}
\usepackage{algorithm}
\usepackage{algorithmic}
\usepackage{varwidth}
\usepackage{parskip}
\usepackage{hyperref}
\usepackage{rotating}
\usepackage[numbers,sort&compress]{natbib}
\usepackage{multirow}
\usepackage{pdflscape}
\usepackage[numbers]{natbib}
\usepackage{caption}
\usepackage{xcolor}
\usepackage{color}
\usepackage{comment}
\usepackage{subcaption}
\newcommand*{\rom}[1]{\expandafter\@\romannumeral #1}

\newcommand{\bea}{\begin{eqnarray}}
	\newcommand{\eea}{\end{eqnarray}}
\newcommand{\bee}{\begin{eqnarray*}}
	\newcommand{\eee}{\end{eqnarray*}}

\begin{document}
\author{ G. P. Singh $^{1}$\footnote{gpsingh@mth.vnit.ac.in }, Romanshu Garg$^{1}$\footnote{romanshugarg18@gmail.com}, Ashutosh Singh$^{2}$\footnote{ashuverse@gmail.com}
\vspace{.2cm}\\
${}^{1}$ Department of Mathematics,\\ Visvesvaraya National Institute of Technology, Nagpur 440010, Maharashtra, India.
\vspace{.3cm}\\
${}^{2}$ Centre for Cosmology, Astrophysics and Space Science (CCASS),\\
GLA University, Mathura 281406, Uttar Pradesh, India}
\date{}
\title{ A generalized $\Lambda$CDM model with parameterized Hubble parameter in particle creation, viscous and $f(R)$ model framework}
\maketitle
\begin{abstract}
In this study, we construct a theoretical framework based on the generalized Hubble parameter form which may arise within the particle creation, viscous and $f(R)$ gravity theory. The Hubble parameter is scrutinized for its compatibility with the observational data relevant to the late-time universe. By using Bayesian statistical techniques based on $\chi^{2}$ minimization method, we determine model parameters's best fit values for the cosmic chronometer and supernovae Pantheon datasets. For the best fit values, the cosmographic and physical parameters are analyzed to understand the cosmic dynamics in model. We also analyze the model section criterion in comparison to the $\Lambda$ cold dark matter model. 
\end{abstract}
\section{Introduction}\label{sec:1}
The observations of astronomical origin indicates that the universe is expanding with an increasing rate \cite{1998AJ....116.1009R,1999ApJ...517..565P,2020A&A...641A...6P}. 
At present, the precise nature of enigmatic fluid (or field) that is speeding up the expansion of cosmos is unclear to a great extent. This fluid (or field) is also referred as the `dark energy' \cite{copeland2006dynamics}. To elucidate the universe's accelerating expansion, the `cosmological constant' $(\Lambda)$ of General relativity model (visualized as the vacuum energy \cite{weinberg1989cosmological}) is widely accepted candidate for dark energy. Although, the cosmological constant appears to fit well with the observational evidences of the universe but this model faces two main problems namely, the coincidence problem and fine tuning problem \cite{copeland2006dynamics}. There is a nearly $120$-order of magnitude difference between its value from particle physics and the value needed to suit cosmic observations \cite{copeland2006dynamics}. 
\vspace{0.3cm}\\
The universe's accelerated expansion may be explained by different cosmological mechanisms, see \cite{capozziello2006cosmological, nojiri2007introduction, nojiri2011unified,odintsov2023recent,bamba2012dark,rev1,rev2,bamba2014inflationary}. A modified gravity theory such as the $f(R)$ gravity, a barotropic fluid model or a combination of both may be used to describe the early-time and late-time accelerated expansion epochs \cite{capozziello2006cosmological, nojiri2007introduction, nojiri2011unified,bamba2012dark,rev1,rev2,bamba2014inflationary}.
An alternative explanation for the accelerated phase can be found in the particle production mechanism also \cite{abramo1996inflationary,zimdahl2000cosmological}. Schrodinger \cite{schro1939proper} proposed the particle generation mechanism which further studied by Parker \cite{parker1968particle, parker1971quantized}. The particle creation mechanism by an external gravitational field in cosmological modeling has been studied by Parker \cite{parker1969quantized}. There has been lots of discussion on the production of matter in an expanding universe \cite{durrer2002particle, singh1999bulk, singh2000particle,singh2020complete, hulke2020variable, singh2002viscous, singh2011anisotropic, chaubey2012bianchi, singh2020study}. The particle creation mechanism may yield cosmological scenarios with non-equilibrium thermodynamical descriptions ranging from inflation to late-time acceleration \cite{chakraborty2014complete}.
\vspace{0.3cm}\\
Viscous fluid presents a compelling and captivating concept within the realms of fluid mechanics \cite{landau2013fluid} and cosmology \cite{misner1967transport, weinberg1971entropy}. Murphy\cite{murphy1973big} has shown that initial singularity in homogeneous and isotropic Friedmann-Lemaitre-Robertson-Walker (FLRW) cosmology can be resolved by continuous bulk viscosity. The effect of viscosity introduction on the formation of singularity in the Friedmann cosmology framework with the idealized assumption of constant bulk viscosity coefficient has been demonstrated by Heller et al.\cite{heller1973imperfect}. Numerous authors have examined the impact of cosmic viscous fluids on universe evolution\cite{singh2016bouncing, singh2018thermodynamical, chaubey2012bianchi, chaubey2016general,garg2024cosmological,singh2021unified,singh2017bulk,singh2017hypersurface,RRAS2020,ASetal2022,ASetal2023,Asepjc2023,sharif2014effects,myrzakulov2015inhomogeneous,AP2024,sym14122630}. $\Lambda$CDM with viscous cosmology was studied in\cite{brevik2017viscous, odintsov2020testing}. The viscous fluids may also pave a way for graceful exit in the early universe during the inflationary epoch \cite{myrzakulov2015inhomogeneous}. The resolution of the initial singularity problem in mainstream cosmology emphasized in the dynamics of dark matter with bulk viscosity effects \cite{szydlowski2020viscous}.
\vspace{0.3cm}\\
In order to comprehend the different aspects of the observable cosmos, various pioneering concepts have been proposed with the modification of general relativity (GR), see for example \cite{capozziello2006cosmological, nojiri2007introduction, nojiri2011unified,bamba2012dark,bamba2014inflationary} and references therein. A fundamental modification originates by replacing the Ricci scalar $(R)$ with a general function of $R$ in the Einstein-Hilbert action of GR, known as $f(R)$ gravity \cite{buchdahl1970non,1982GReGr..14..453K}. $\Lambda$CDM bounds for $F(R)$ gravity were studied in\cite{nojiri2007unifying, odintsov2023exponential}. The $f(R)$ gravity provides a good explanation for the cosmological expansion phenomenon and several studies have examined the limitations of feasible cosmological theories \cite{capozziello2006cosmological,amendola2007f}. A number of authors\cite{starobinsky2007disappearing, capozziello2008solar, liu2018constraining} have discussed observational deductions from the models of $f(R)$ and, have talked about the solar system's constraints as well as the astrophysical phenomenon in $f(R)$ gravity \cite{capozziello2008cosmography, singh2020cosmological}.
\vspace{0.1cm}\\
In this paper, we show that a parameterized form of Hubble parameter may be a solution in particle creation model. Independently, the considered Hubble parameter may also solved in bulk viscous model and the $f(R)$ gravity model. In cosmological modeling, the ansatzes of different cosmological quantities are studied to check whether they admit physically reasonable behaviors or not? The observational viability of ansatzes subjected to the observational data such as the Cosmic chronometer data and Pantheon supernovae data may aid in ruling out different kind of parametrizations. In this paper, we proceed along these lines and show that the cosmological solution based on parameterized Hubble parameter may be physically admitted.
\vspace{0.3cm}\\
The paper has been arranged in $6$ sections as follows: In Sec.(\ref{sec:2}), the characteristics of cosmological solutions are shown by using a particular Hubble function which may explain the transition from early era deceleration into late-time acceleration. In Section (\ref{sec:3}), the compatibility of cosmological solution is investigated by using Bayesian statistical techniques with two observational datasets, namely the cosmic chronometer (CC) and Pantheon datasets. In Sec.(\ref{sec:4}), we show that the considered Hubble parameter may be a solution in cosmological frameworks such as particle creation, viscous fluid model and $f(R)$ gravity model. We construct the $f(R)$ gravity form and examine the dynamical characteristics of universe in the model. In Sec.(\ref{sec:5}), we discuss the cosmographic parameters along with the age of current universe in model. Finally, the summary and conclusions are given in Sec.(\ref{sec:7}).

\section{The cosmological equations and background dynamics}\label{sec:2}
The astronomical observations about the expanding universe points that the rate of universe's expansion is increasing although its spatial geometry seems be almost flat \cite{2020A&A...641A...6P}. It is also well-established that the universe expansion history has a past of decelerating expansion which transits into the accelerating expansion phase during present times. For the observable universe, as par Aghanim et al. \cite{2020A&A...641A...6P} the present day value of the expansion rate is $H_0=67.4\pm 0.5 \ km/(s\cdot Mpc)$. Homogeneous and isotropic universe's expansion rate is described by $H=\frac{\dot{a}}{a}$, where $a$ denotes the scale factor and overhead dot is time derivative. The spatially flat metric for homogeneous and isotropic universe may be written as
\begin{equation}
	ds^2=-dt^2+a^2(dr^2+r^2(d\theta^2+\sin^2\theta d\phi^2))
	\label{eq1a}
\end{equation}
In the general relativity context, the field equations can be expressed as
\begin{equation}
	3H^2=k^2\rho, \quad 2\dot{H}+3H^2=-k^2p 
	\label{eq2a}
\end{equation}
where $\rho$ and $p$ represent the energy density and pressure respectively. In order to understand the transitional universe expansion era, we adopt a parametrization of the Hubble parameter $H(t)$. The parametrizations may be used in a model-independent manner for exploring the characteristics of dark energy dominated universe. Different kind of parameterization when tested against the observational data may single out the physically reasonable cases. In this approach, we proceed with the Hubble parameter having form 
\begin{equation}{\label{1}}
H(a)=H_{0}\sqrt{\frac{m}{a^{n}}+1-m}   
\end{equation}
Here, $H_{0}$ represents the Hubble parameter's current value and $m$, $n$ are arbitrary constants. The above Hubble parameter may be seen as the generalization of $\Lambda$ cold dark matter model defined by $H(a)=H_{0}\sqrt{\frac{\Omega_m}{a^{3}}+\Omega_\Lambda}$. Although, the $\Lambda$CDM model is supported by the latest observational datasets \cite{1998AJ....116.1009R, 2020A&A...641A...6P,perlmutter1998discovery,eisenstein1998cosmic,haridasu2017strong}, the cosmological studies based on parametrizations of different parameters are widely explored \cite{bamba2014inflationary,singh1999bulk,singh2000particle,
singh2020complete,hulke2020variable,
singh2002viscous,singh2011anisotropic,
chaubey2012bianchi,singh2020study,
chakraborty2014complete,singh2016bouncing,
singh2018thermodynamical,chaubey2016general,
garg2024cosmological,singh2021unified,singh2017bulk,
singh2017hypersurface,
RRAS2020,ASetal2022,ASetal2023,Asepjc2023,sharif2014effects,
myrzakulov2015inhomogeneous,M2020,Smetal2022,
Lalkeetal2023,AsSk2024,M2024,reva1, reva2}. Above equation (\ref{1}) may be solved using the relation $a=\frac{a_{0}}{1+z}$ with standard convention $a_{0}=1$ for present era to obtain Hubble parameter in terms of red-shift ($z$) as
\begin{equation}{\label{3}}
H(z)=H_{0}\sqrt{m(1+z)^{n}+1-m}   
\end{equation}
This kind of Hubble parameter may arise with affine equation of state $p=n_1\rho-n_2$, where $n_i,i=1,2$ are some constant \cite{AsSk2024} or in modified theory of gravity \cite{M2020,M2024}.
When it comes to understanding the dynamics of the cosmic universe the cosmographical parameter\cite{rev2} with the most influential is thought to be $q$. Based on the values of this parameter, the expansion dynamics of universe may be categorized into either accelerated or decelerated phases with $q=0$ representing the transitional periods. The universe exhibits deceleration when $q>0$ and power-law expansion when $-1<q<0$. The universe exhibits super-exponential (de Sitter) expanding era for $q<-1$ ($q=-1$) respectively. This cosmographic parameter may be defined as $q=-1+\frac{d}{dt}\frac{1}{H}$ and by using equation (\ref{3}), we obtain 
\begin{equation}{\label{5}}
q=-1+\frac{mn(1+z)^{n}}{2\left[m(1+z)^{n}+1-m\right]}
\end{equation}
\begin{figure}[!htb]
\captionsetup{skip=0.4\baselineskip,size=footnotesize}
   \begin{minipage}{0.40\textwidth}
     \centering
     \includegraphics[width=8cm,height=5.5cm]{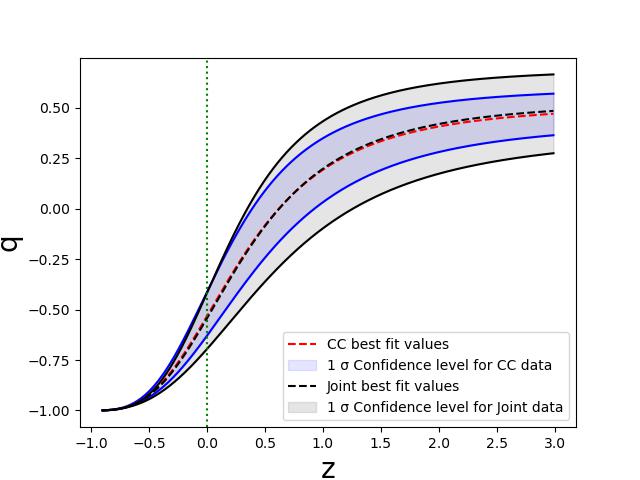}
\caption{$\mathit{q}$ with $\mathit{z}$}
\label{fig:1}
    \end{minipage}\hfill
   \begin{minipage}{0.45\textwidth}
     \centering
     \includegraphics[width=8cm,height=5.5cm]{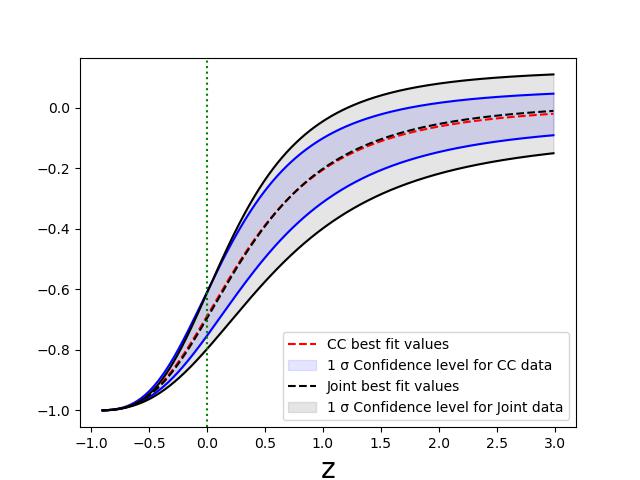}
  \caption{ $\omega_{eff}$ with $\mathit{z}$ }
\label{fig:2}
   \end{minipage}
\end{figure}
Based on the model parameters' best fit values (see section \ref{sec:3}), the evolution of deceleration parameter and effective equation of state (EoS) parameter $\left( \text{defined as} \ \omega_{eff}=-1-\frac{2}{3}\frac{\dot{H}}{H^2}\right) $ of the reconstructed universe is shown in figure (\ref{fig:1}) and (\ref{fig:2}) respectively. For our model, the values of deceleration parameter are $q_{0}=-0.533$ (for the CC data) and $q_{0}=-0.5441$ (for the joint data) at present time having $z=0$. This $q_{0}$ value is very close to $q_{0}=-0.55$ for the $\Lambda$CDM model\cite{rev2}.
\section{Observational constraints on model}\label{sec:3}
Here, we examine the compliance of parameterized from of Hubble parameter (\ref{1}) with datasets of the cosmic chronometer (CC) and the joint data consisting of Pantheon and cosmic chronometer data nomenclated (in present paper) as CC+Pantheon sample. In order to perform statistical analysis, we use $ \chi^{2}$ minimization method with the Markov Chain Monte Carlo (MCMC) technique implemented with the emcee tool \cite{foreman2013emcee} and constrain the parameters to study of the cosmic behavior in the model.
\subsection{The Cosmic chronometer data}\label{subsec:3.1} 
\begin{figure}[!htb]
\captionsetup{skip=0.4\baselineskip,size=footnotesize}
\centering
\includegraphics[width=9cm,height=7cm]{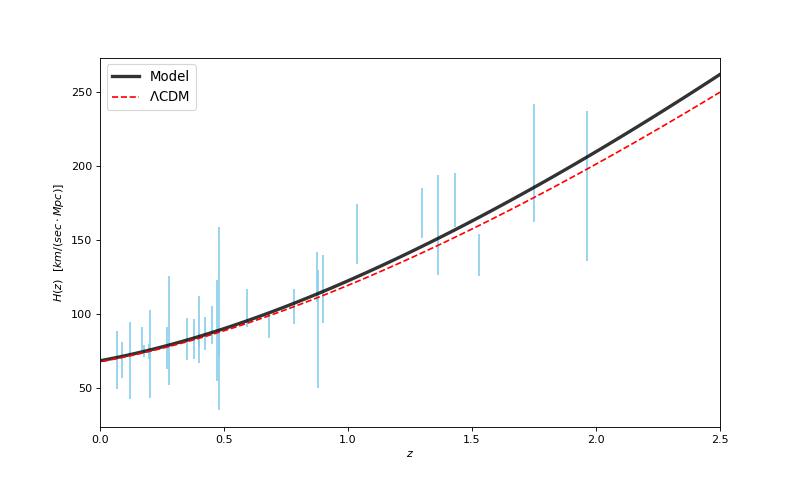}
\caption{ The best fit $H(z)$ curve with $\mathit{z}$ for the present model as compared to the $\Lambda$CDM  model}
\label{fig:4}
\end{figure}

The values of Hubble parameter at any instant describes the expansion rate of the universe at that particular instant and its observational values are important for exploration of dark energy and the evolution of universe. To constrain the model parameters, we use Cosmic Chronometer data composed of 31 data points \cite{sharov2018predictions, simon2005constraints} which are determined by differential ages of galaxies technique within the red-shift region $0.07 \leq z \leq 1.965$.\\
In this case, we minimize the $\chi^{2}$ function and thus the model parameters $H_{0}, m,$ and $n$ are estimated with their median values. The corresponding $\chi^2$ may be expressed as 
\begin{equation}{\label{6}}
\chi^{2}_{CC}(\theta)=\sum_{i=1}^{31} \frac{[H_{th}(\theta,z_{i})-H_{obs}(z_{i})]^{2}    }{ \sigma^{2}_{H(z_{i})}}   
\end{equation} 
where the observed values are denoted by $H_{obs}(z_{i})$ and theoretical values of the Hubble parameter are denoted by $H_{th}(\theta , z_{i})$. The values $\sigma^{2}_{H(z_{i})}$ are the standard deviation for each $H_{obs}(z_{i})$ observed value. \\
Figure $(\ref{fig:4})$ illustrates the error bars of the CC points with the best fit Hubble parameter curve for the Hubble parameter (Eq. (\ref{3})).

\subsection{The Pantheon data}\label{subsec:3.2}
\begin{figure}
	\includegraphics[width=13cm,height=13cm]{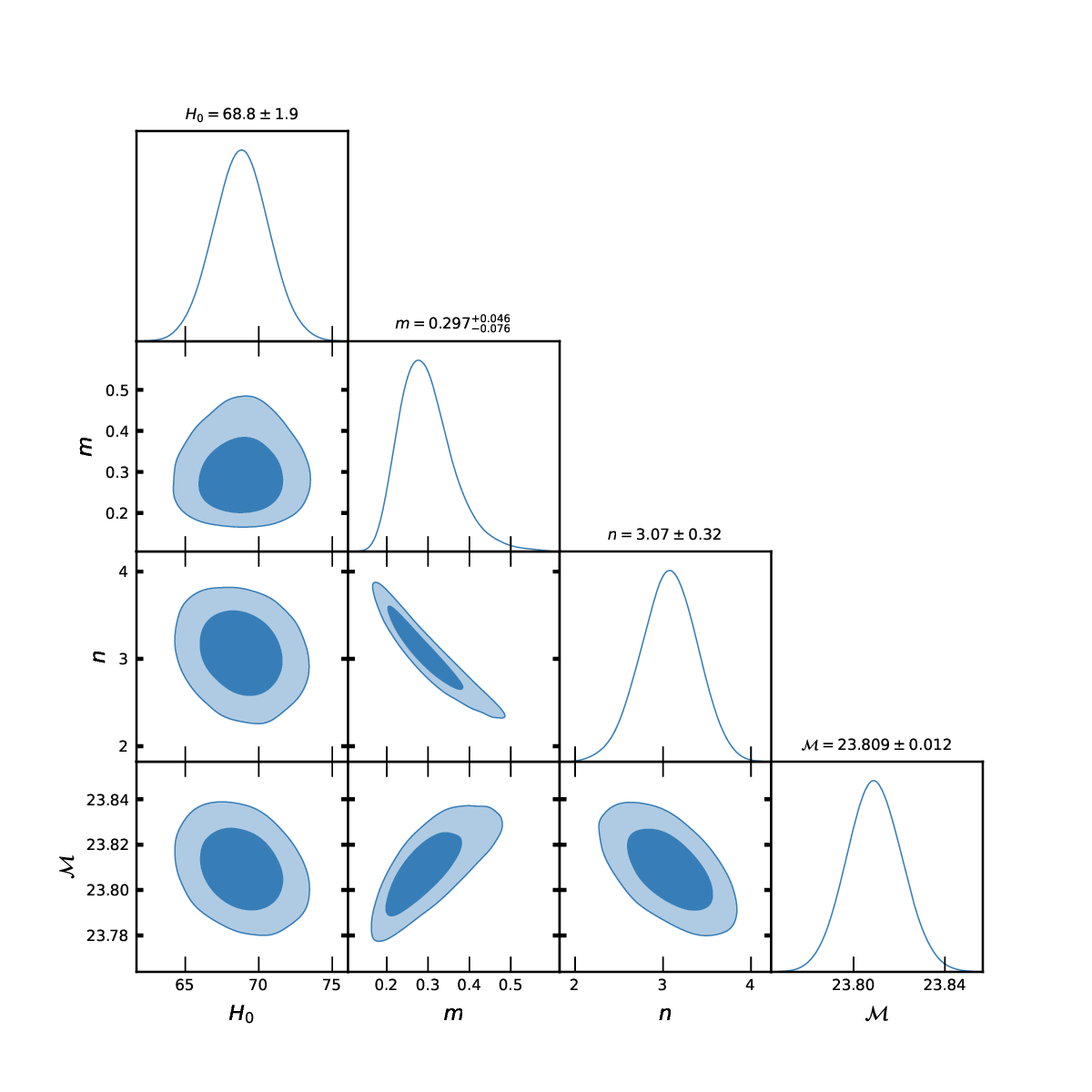}
	\caption{1$\sigma$ and $ 2\sigma$ marginalized contour map and $1$D posterior distributions for $H_{0}, \mathit{m},\mathit{n}$ and $\mathcal{M}$ using the Joint CC+Pantheon data set.}
	\label{fig:5}
\end{figure}

We use the Pantheon sample, which includes $1048$ supernovae Type Ia (SNIa) data points for the red-shift range $0.01 < z < 2.26$ \cite{scolnic2018complete}. The CfA1-CfA4 \cite{riess1999bvri, hicken2009improved} surveys, Pan-STARRS1 Medium Deep Survey \cite{scolnic2018complete}, SDSS \cite{sako2018data}, SNLS \cite{guy2010supernova}, Carnegie Supernova Project (CSP) \cite{contreras2010carnegie} contributes to the SNIa sample. For the MCMC analysis using Pantheon data, the theoretically expected apparent magnitude $\mu_{th}(z)$ is given by
\begin{equation}{\label{7}}
\mu_{th}(z)=25+5\log_{10}\left[\frac{d_{L}(z)}{Mpc}\right]+M
\end{equation}
where $M$ is the absolute magnitude. Also, the luminosity distance $d_{L}(z)$ (having dimension of the \textit{Length}) may be defined as \cite{odintsov2018cosmological}
\begin{equation}{\label{8}}
d_{L}(z)=c(1+z)\int_{0}^{z}
\frac{dz'}{H(z')}
\end{equation}
where $z$ represents SNIa's red-shift as determined in the cosmic microwave background (CMB) rest frame and and $c$ is the speed of light. The luminosity distance $(d_{L})$ is typically substituted with the dimensionless Hubble-free luminosity distance given by $D_{L}(z) \equiv H_{0}d_{L}(z)/c$. The equation (\ref{7}) could also be rewritten as
\begin{equation}{\label{9}}
\mu_{th}(z)=25+5\log_{10}\left[\frac{c/H_{0}}{Mpc}\right]+M+5\log_{10}\left[D_{L}(z)\right] 
\end{equation}
The parameters $M$ and $H_{0}$ can be combined to create a new parameter $\mathcal{M}$, which may be identified as
\begin{equation}{\label{10}}
\mathcal{M}\equiv  M+25+5\log_{10} \left[\frac{c/H_{0}}{Mpc}\right]=M+42.38 -5\log_{10}(h)
\end{equation}
where $H_{0}=h \times 100 \ Km/(s\cdot Mpc)$. We use this parameter with pertinent $\chi^{2}$ for Pantheon data in the MCMC analysis as \cite{asvesta2022observational}
\begin{equation}{\label{11}}
\chi^{2}_{P}= \nabla \mu_{i}C^{-1}_{ij}\nabla \mu_{j}
\end{equation}
where $\nabla \mu_{i}=\mu_{obs}(z_{i})-\mu_{th}(z_{i})$, $C_{ij}^{-1}$ is the covariance matrix's inverse and $\mu_{th}$ will be provided by equation (\ref{9}). The luminosity distance depends on the Hubble parameter. Therefore, we use the emcee package \cite{foreman2013emcee} and equation (\ref{3}) to get the maximum likelihood estimate using the joint CC+Pantheon data set. The joint $\chi^{2}$ for maximum likelihood estimate may be defined as $\chi^{2}_{CC}+\chi^{2}_{P}$. In Fig. $(\ref{fig:5})$, we displays the posterior distribution with $1\sigma$ and $2\sigma$ from the Monte Chain Monte Carlo analysis using CC+Pantheon data. The Table (\ref{table:1}) summarizes the best fit values for the model parameters in the MCMC analysis for the model.
\begin{table}[htbp]
\centering
\begin{tabular}{|c|c|c|c|c|}
\hline
Dataset & $H_{0} \ [Km/(s\cdot Mpc)]$ & m & n & $\mathcal{M}$ \\
\hline
CC & $68.326^{+1.005}_{-1.045}$ & $0.307^{+0.059}_{-0.050}$ &$3.040^{+0.165}_{-0.164}$ & - \\
\hline
CC+Pantheon  & $68.8^{+1.9}_{-1.9}$  & $0.297^{+0.046}_{-0.076}$  &  $3.07^{+0.32}_{-0.32}$ & $23.809^{+0.012}_{-0.012}$\\
\hline
\end{tabular}
\caption{The model parameters and their median values in MCMC analysis}
\label{table:1}
\end{table}

\section{The parameterized Hubble parameter as cosmological solution in different frameworks }\label{sec:4}
\subsection{The Particle Creation model}\label{subsec:4.1}
\begin{figure}[!htb]
\captionsetup{skip=0.4\baselineskip,size=footnotesize}
   \begin{minipage}{0.40\textwidth}
     \centering
     \includegraphics[width=8.5cm,height=5.5cm]{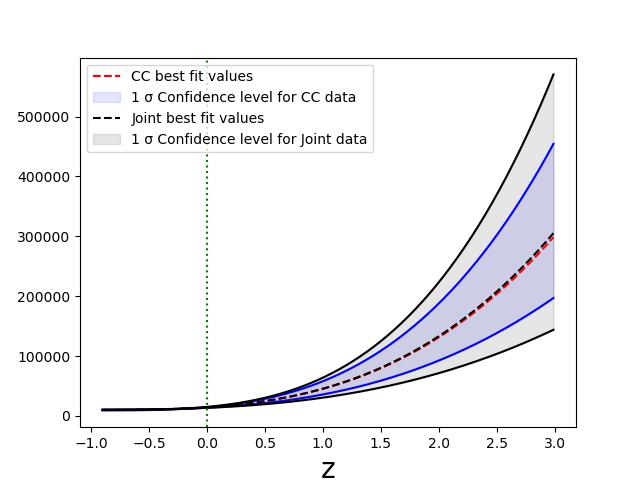}
\caption{Energy density $(\rho_{m})$ with $\mathit{z}$}
\label{fig:6}
    \end{minipage}\hfill
   \begin{minipage}{0.40\textwidth}
     \centering
     \includegraphics[width=8.5cm,height=5.5cm]{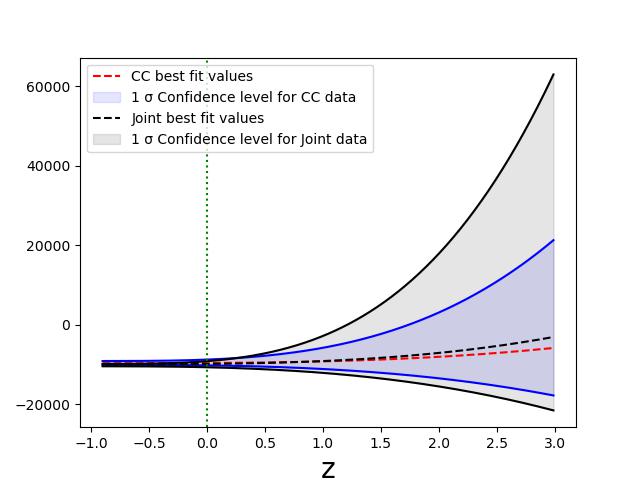}
  \caption{Creation pressure ($\mathit{p_{c}}$)  with $\mathit{z}$}
\label{fig:7}
   \end{minipage}
\end{figure}

With the FLRW spacetime, the fundamental cosmological equations with the particle creation mechanism in a model may be written as	
\begin{equation}{\label{13}}
3H^{2}=k^{2}\rho, \quad 2\dot{H}+3H^{2}=-k^{2}(p+p_{c})
\end{equation}
where $p$ and $\rho$ represent to the thermodynamic pressure of the matter content and energy density respectively with $p_{c}$ is representing the creation pressure. The fluid particles are not conserved if thermodynamic system is regarded as open ($N^{i}_{;i}=n\Gamma \neq 0$, where particle number density is represented by $n$, while the fluid flow vector is denoted by $N^{i}=nu^{i}$). The particle conservation equation can be expressed as follows, 
\begin{equation}{\label{15}}
\dot{n}+3nH=n\Gamma
\end{equation}
where the particle production rate $\Gamma $ is the rate of change in particle number in the co-moving volume $V$. Depending on the particle production rate, one may identify the creation or annihilation of particles. Particle creation occurs when $\Gamma > 0$, particle annihilation occurs when $\Gamma < 0 $, and no particle production occurs when $\Gamma = 0 $. According to the second law of thermodynamics, $\Gamma > 0$ is required for entropy to never decrease. The gravitationally induced adiabatic particle formation rate and creation pressure $p_{c}$ are related by the  
relation\cite{calvao1992thermodynamics, lima1992equivalence}.
\begin{equation}{\label{16}}
p_{c}=\frac{-\Gamma}{3H}(\rho +p)
\end{equation}
When particle production is present or absent, the creation pressure $p_{c}$ is  negative or zero respectively. The energy conservation equation may be expressed as \begin{equation}{\label{17}}
\dot{\rho}+3H(\rho +p)=-3Hp_{c}
\end{equation}
By considering the universe with matter having $p_m=0$, the energy density of matter may be written using Eqs. (\ref{3}), (\ref{13}) and (\ref{17}) as
\begin{equation}{\label{18}}
\rho_{m}=3H^{2}, \quad p_{c}=\frac{HH_{0}mn(1+z)^{n}}{\sqrt{m(1+z)^{n}+1-m}}-3H^{2}
\end{equation}
In this scenario, the particle creation rate may be written as using Eqs. (\ref{3}), (\ref{13}), (\ref{16}) as
\begin{equation}
\Gamma=\left( 3-n\right)H-\frac{H_0^2n(m-1)}{H}
\end{equation}
The above derived particle creation rate may interpolate between the early decelerating era with the late-time accelerating era in the model. The creation pressure being negative from the recent past may yield the mechanism for accelerating universe expansion in model. Matter's energy density behavior and creation pressure have been displayed in Fig. (\ref{fig:6}) and (\ref{fig:7}) respectively. The energy density remains positive during different cosmological evolution eras while the creation pressure has become negative from the recent past.

\subsection{Bulk viscous model}\label{subsec:4.2}
\begin{figure}[!htb]
\captionsetup{skip=0.4\baselineskip,size=footnotesize}
   \begin{minipage}{0.40\textwidth}
     \centering
     \includegraphics[width=8.5cm,height=5.5cm]{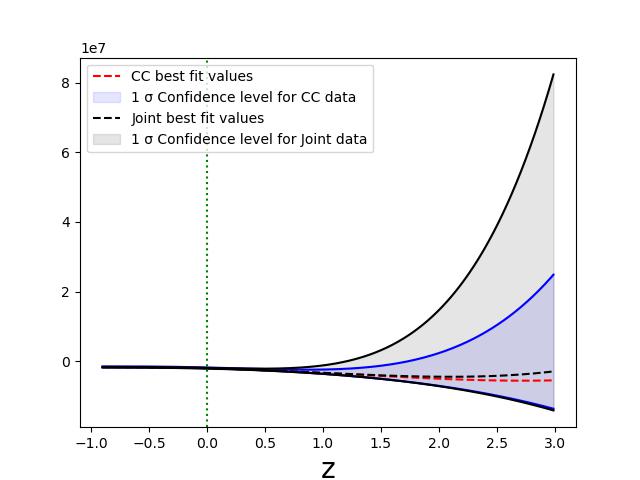}
\caption{Bulk viscous pressure $\mathit{(p')}$ with $\mathit{z}$ }
\label{fig:8}
    \end{minipage}\hfill
   \begin{minipage}{0.40\textwidth}
     \centering
     \includegraphics[width=8.5cm,height=5.5cm]{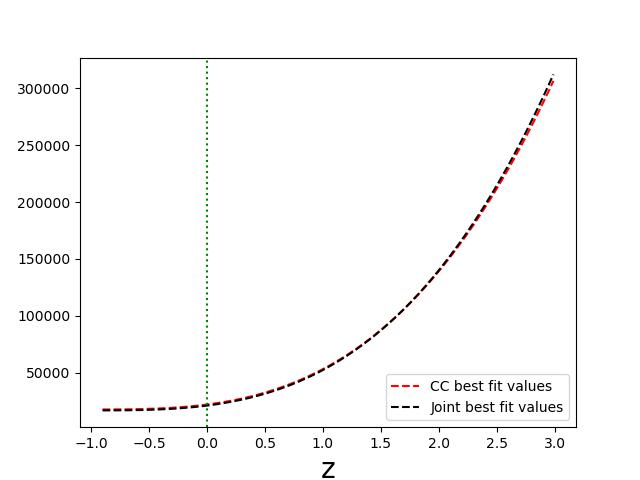}
  \caption{Energy density $(\rho_{m})$ with $\mathit{z}$ for $f(R)$ gravity model}
\label{fig:9}
   \end{minipage}
\end{figure}

For the FRW metric (\ref{eq1a}), the field equations in the bulk viscous fluid model are 
\begin{equation}{\label{20}}
3H^{2}=k^{2}\rho, \quad 2\dot{H}+3H^{2}=-k^{2}p'
\end{equation}
where the bulk viscous pressure $p'$ may be written as 
\begin{equation}{\label{22}}
p'=p-3H\xi
\end{equation}
with $\xi$ being coefficient of bulk viscosity with $\rho$ and $p$ denoting the energy density and pressure of fluid respectively. In this paper, we consider the generic form of the inhomogeneous viscous fluid having form $p'=p -3H\xi(a, H, \dot{H},....)$, where the bulk viscosity $\xi(a, H, \dot{H},....)$ is a general function of $a$, $H$ and it’s derivatives. The continuity equation may be expressed as 
\begin{equation}{\label{23}}
\dot{\rho}+3H(\rho +p)=3H\xi
\end{equation}
where $\xi$ is a natural candidate for actual fluid and contributes to the dissipative effects. By considering the matter having $p_m=0$, the ernergy density and viscous pressure may take the form  
\begin{equation}{\label{24}}
\rho_{m}=3H^{2}, \quad \xi= (3-n)H^2+{H_0}^2n(1-m)
\end{equation}
The Fig. $(\ref{fig:8})$ depict the evolution of bulk viscous pressure $(p')$ with red-shift. We can easily observe that bulk viscous pressure $(p')$ has negative values which may be responsible for accelerating universe expansion in the model.

\subsection{The $f(R)$ gravity model reconstruction}\label{subsec:4.3}
The $f(R)$ gravity theory action may be written as\cite{rev1,rev2}
\begin{equation}{\label{27}}
s=\int d^{4}x \sqrt{-g}f(R)+S_{m}
\end{equation}
where $f(R)$ denotes a general function of the Ricci scalar $(R)$ and the action for the appropriate matter distribution is denoted by $S_{m}$. The field equation may result by varying action (\ref{27}) with respect to the metric tensor  $g_{ij}$ as
\begin{equation}{\label{28}}
f'R_{\mu v}-\frac{f}{2}g_{\mu v}-(\nabla_{\mu}\nabla_{v}-g_{\mu v}\Box)f'=T_{\mu v}
\end{equation}
where $f'(R)=\frac{\partial f(R)}{\partial R}$ and $T_{\mu v}=\frac{-2}{\sqrt{-g}}\frac{\delta S_{m}}{\delta g^{ij}}$ is stress-energy tensor. The field equations of $f(R)$ gravity can be expressed in terms of the Einstein tensor, which includes an effective energy–momentum tensor $T^{eff}_{\mu v}$.  We may write
\begin{equation}{\label{26}}
G_{\mu v}=\frac{1}{f'(R)}(T_{\mu v}+T_{\mu v}^{eff})
\end{equation}
The Raychaudhuri equation in $f(R)$ gravity for a time-like congruence with velocity vector $u^{i}$ may take the form \cite{raychaudhuri1955relativistic, ehlers1993contributions}
\begin{equation}{\label{29}}
\frac{d \Theta}{d \tau}=-\frac{\Theta^{2}}{3}+\nabla_{i}a^{i}-\sigma_{ij}\sigma^{ij}+\omega_{ij}\omega^{ij}-R_{ij}u^{i}u^{j}
\end{equation}
The last term that is located on the right-hand side of equation (\ref{29}) can be expressed as follows using the field equations (\ref{28}) for $f(R)$ theory:
\begin{equation}{\label{30}}
R_{ij}u^{i}u^{j}=\frac{1}{f'}\left[T_{ij}+\frac{f}{2}g_{ij}+(\nabla_{i}\nabla_{j}-g_{ij}\Box)f'\right]  u^{i}u^{j}
\end{equation}
In this paper, we are dealing with the homogeneous and isotropic, spatially flat universe given by the metric (\ref{eq1a}). The Raychaudhuri equation (\ref{29}) takes the following form for such a metric and a matter distribution of a perfect fluid $T^{ij}=(\rho +p)u^{i}u^{j}+pg^{ij}$ as \cite{guarnizo2011geodesic,gupta2019reconstruction}
\begin{equation}{\label{31}}
\frac{\ddot{a}}{a}=\frac{1}{f'}\left( \frac{f}{6}+Hf''\dot{R}-\frac{\rho}{3}\right)
\end{equation}
It should be emphasized that for the fluid distribution, we have not assumed any equation of state until now,  but by using field equations (\ref{26}) in the Raychaudhuri equation (\ref{29}), one may eliminates the fluid pressure $p$.\\
In principle, the $f(R)$ function of this modified gravity may be determined by using either the Hubble parameter or the scale factor. Reconstruction of $f(R)$ for different Hubble parameter was initiated in Ref. \cite{nojiri2009cosmological}. The late- and early-time acceleration may be realized in $f(R)$ models with reconstruction approach where dark energy may have the quintessence-like behavior \cite{rev1}. A detailed summary of $f(R)$ gravity reconstruction in metric formalism as well as Palatini formalism with the cosmological viability conditions has been presented in Ref. \cite{rev2}. Here, we use Eq. (\ref{1}) with Eq. (\ref{31}) to determine $f(R)$ gravity form. The Ricci scalar $R$ for a spatially flat Friedmann-Robertson-Walker metric (\ref{eq1a}) is $R=6\left(\frac{\dot{a}^{2}}{a^{2}}+\frac{\ddot{a}}{a}\right)$. Using the equation (\ref{1}), we can write
\begin{equation}{\label{32}}
R=\alpha.a^{-n}+\beta
\end{equation}
Here $\alpha=(12H_{0}^{2}m-3H_{0}^{2}mn)$ and $\beta =12H^{2}_{0}(1-m)$. Using the equation (\ref{1}), (\ref{31}) and (\ref{32}), we obtain
\begin{eqnarray}{\label{33}}
-n \alpha \left(\frac{R-\beta}{\alpha}\right)\left[H^{2}_{0}m\left(\frac{R-\beta}{\alpha}\right)+H^{2}_{0}(1-m)\right]f''(R)+\left[\left(\frac{R-\beta}{\alpha}\right)\left(\frac{H_{0}^{2}mn}{2}-H^{2}_{0}m\right)-(1-m)H^{2}_{0}\right]f'(R)
 \nonumber \\
 +\frac{f}{6}=\frac{\rho}{3}
\end{eqnarray}
For the matter having pressure $p=0$, the energy density may be given by $\rho=\frac{\rho_{m0}}{a^{3}}$ and thus above Eq. (\ref{33}) may take the form \begin{eqnarray}{\label{34}}
-n \alpha \left(\frac{R-\beta}{\alpha}\right)\left[H^{2}_{0}m\left(\frac{R-\beta}{\alpha}\right)+H^{2}_{0}(1-m)\right]f''(R)
 +\left[\left(\frac{R-\beta}{\alpha}\right)\left(\frac{H_{0}^{2}mn}{2}-H^{2}_{0}m\right)-(1-m)H^{2}_{0}\right]f'(R)
 \nonumber \\ 
 +\frac{f}{6}=\frac{\rho_{m0}}{3}\left(\frac{R-\beta}{\alpha}\right)^{\frac{3}{n}}
\end{eqnarray}
Solving Eq. (\ref{34}), we obtained one solution for $f(R)$ as
\begin{equation}{\label{35}}
f(R)=-6\frac{\rho_{m0}}{\alpha}\left(\frac{\beta}{3}+I\right)+\frac{\rho_{m0}}{\alpha}R
\end{equation}
 where $I=(m-1)H^{2}_{0}- \frac{H^{2}_{0}m\beta(\frac{n}{2}-1)}{\alpha}$, $\alpha=(12H_{0}^{2}m-3H_{0}^{2}mn)$ and $\beta =12H^{2}_{0}(1-m)$. We may also have the energy density of the matter given by 
\begin{equation}{\label{38}}
\rho_{m}=3H_{0}^{2}\left[m(1+z)^{n}+1-m\right]+\beta+3I
\end{equation}
For the median values given in Table (\ref{table:1}), the behavior of energy density with red-shift have been displayed in Fig. (\ref{fig:9}). The energy density remains positive during the decelerating era and will preserve its positive nature as the universe evolves into the accelerating era in model. The energy density of matter (Eq. (\ref{38})) in the terms of red-shift decreases as the time evolves in the expanding model based on the expansion rate (\ref{3}). The parameter involved in the expression of $\rho_{m}$ describes the evolution of matter density in accord to the observations, in which the energy density of matter is decreasing from the past.\\
The present model describing early deceleration transiting into late-time acceleration have the $f(R)$ form $f(R)=a_1R+a_2$, where $a_1,a_2$ are recombined constants. The $f(R)$ gravity may also have cosmologically viable models describing early era inflation as well as the late era accelerated expansion \cite{rev3}. The power-law $f(R)$ cosmology may describe the inflationary expansion and may relate with the $\Lambda$CDM dynamics in a natural way \cite{rev4}. The non-singular exponential $f(R)$ cosmology may also relate the early era and late era accelerated universe evolution \cite{rev5}. In this sense, the present $f(R)$ model may belong to the power-law $f(R)$ model and it may relate the early deceleration to the late-time acceleration.

\section{General issues}\label{sec:5}
In this section, we study the cosmographic evolution of the universe governed by Hubble parameter (\ref{1}). The cosmographic parameter helps to identify the sharp contrast between the dark energy model and the $\Lambda$CDM model. We also identify the universe's age in model for the best fit values in model.
\subsection{Cosmographic parameters}\label{sec:5.1}
The kinematic quantities like as jerk $(j)$ and snap $(s)$ parameter defines the cosmic scenario by using the geometric quantities like as the scale factor$(a)$ and its derivatives. The cosmography study was initially discussed by Weinberg \cite{weinberg2008cosmology} by using a Taylor series to introduce the scale factor that increased around present time $t_{0}$. The Hubble parameter $H$ is regarded as a varying observable quantity prior to discovering evidence of the Universe's accelerating expansion. The deceleration parameter $q$ illustrates its evolution, by using the second-order derivative of $a$ \cite{mukherjee2016parametric}. The snap $(s)$ and jerk $(j)$ parameters provide  insights about the cosmic evolution of the universe. Jerk can also sometimes be referred to as jolt, and jounce is another word for snap and may be defined as \cite{rev2,visser2004jerk}:
\begin{equation}{\label{39}}
\mathit{j}=\frac{1}{aH^{3}}\left(\frac{d^{3}a}{dt^{3}}\right),\ \ \mathit{s}=\frac{1}{aH^{4}}\left(\frac{d^{4}a}{dt^{4}}\right)
\end{equation}
The cosmographic analysis in absence of spatial curvature may be well understood with the expression (\ref{39}). However, the presence of spatial curvature in a model leads to limitation in the standard cosmographic approach \cite{rev2}. The standard cosmography may be improved with the method of Padé polynomials \cite{rev6,rev7,rev8,rev9,rev10}, the method of Chebyshev polynomials \cite{rev8a,rev2}, for more details see, Ref. \cite{rev2,rev11,rev12,rev13}. For $\mathit{j}$ and $\mathit{s}$ in present analysis the equation (\ref{39}) can be rewritten in red-shift's terms as \cite{wang2009probing}. 
\begin{equation}{\label{40}}
j(z)=\left(1+2q(z)\right)q(z)+\frac{dq}{dz}(1+z), \  \ \  \   s(z)=-\left(2+3q(z)\right)j(z) -\frac{dj}{dz}(1+z)
\end{equation}
\begin{figure}[!htb]
\captionsetup{skip=0.4\baselineskip,size=footnotesize}
   \begin{minipage}{0.5\textwidth}
     \centering
     \includegraphics[width=7.5cm,height=5.5cm]{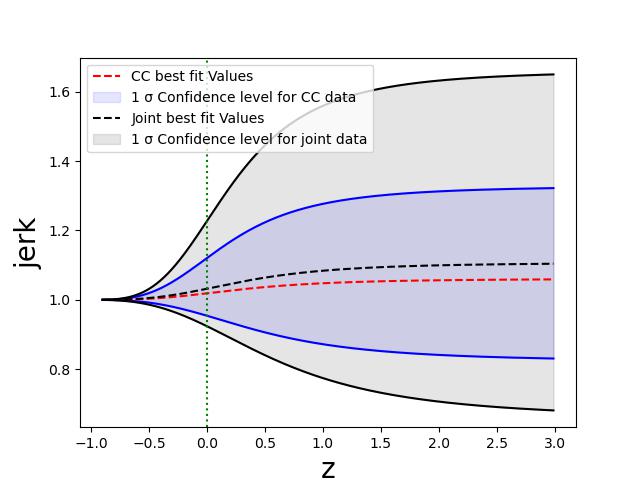}
\caption{Jerk parameter with $\mathit{z}$}
\label{fig:10}
    \end{minipage}\hfill
   \begin{minipage}{0.5\textwidth}
     \centering
     \includegraphics[width=7.5cm,height=5.5cm]{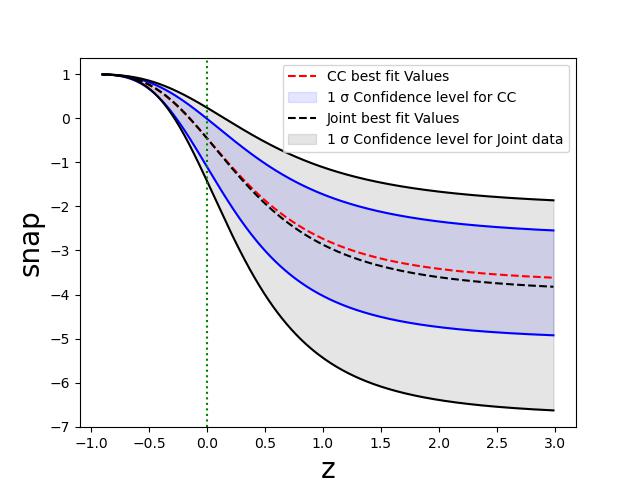}
  \caption{Snap parameter with $\mathit{z}$}
\label{fig:11}
   \end{minipage}
\end{figure}
For the median values, the jerk and snap parameter behaviors have been illustrated in figure $(\ref{fig:10})$ and $(\ref{fig:11})$ respectively. The snap and jerk parameters are provided by $s_{0}=-0.4467$ and $j_{0}=1.0186$ respectively, for the median values derived from the CC data set. The values of the jerk parameter differ from the $\Lambda$CDM model with $j_{0}=1$. The obtained values of jerk and snap parameters respectively are $j_{0}=1.0319$, $s_{0}=-0.4482$ for the median values derived from for joint data set.\\
In the model independent reconstruction of $f(R)$ gravity using Padé approximation \cite{rev8}, $j_0=0.593_{-0.210}^{+0.216}$. However, for third order Taylor's approximation $j_0=1.223_{-0.664}^{+0.644}$, $s_0=0.394^{+1.335}_{-0.731}$ \cite{rev2}. In case of Chebyshev polynomial method, $j_0=1.585^{+0.497}_{-0.914}$, $s_0=1.041^{+1.183}_{-1.784}$ \cite{rev8a}. For the $\Lambda$CDM model with $\Omega_m=0.3$, $q_0=-0.55,j_0=1,s_0=-0.35$ \cite{rev2}. The jerk and snap parameter values of present model are consistent with these findings.
\subsection{Age of the Universe}
In a cosmological model, the cosmic age $t(z)$ of the universe can be calculated as
\begin{equation}{\label{42}}
t(z)= \int_{z}^{\infty} \frac{dz}{(1+z)H(z)} \,dz 
\end{equation}
We numerically calculate the aforementioned integral and get the present age of the universe for $z=0$ and use the Hubble parameter $H(z)$ (Eq. \ref{1}). For this model, the obtained value of universe's present age is $t_{0}=13.52^{+1.73}_{-1.5}$ Gyr for CC data and $t_0=13.4^{+3.30}_{-2.04}$ Gyr  which are very close to the present age values obtained from the recent observations \cite{2020A&A...641A...6P}.
\subsection{Information criteria}
In order to test the statistical performance of model (\ref{3}), we use the popular information criteria named as Akaike information criteria (AIC) and Bayesian information criteria (BIC). The expression for AIC is given by \cite{r11,r12,r13}
\begin{equation}
	\text{AIC}\equiv -2\ln \mathcal{L}_{max}+2p=\chi_{min}^2+2p
	\label{eqr1} 
\end{equation} 
where $p$ is total number of free (fitted) parameters used in the present MCMC analysis. $\mathcal{L}_{max}$ is the maximum likelihood of the considered model. \\
The expression for BIC is given by \cite{r11,r12,r14}
\begin{equation}
	\text{BIC}\equiv -2\ln \mathcal{L}_{max}+p\ln N
	\label{eqr2} 
\end{equation} 
where $p$ is total number of free (fitted) parameters used in the present MCMC analysis. Using the definitions of AIC and BIC, we calculate the $\triangle \text{AIC}$ and $\triangle \text{BIC}$ as compared to $\Lambda$CDM model. According to the criterion of calibrated Jeffrey's scale \cite{r15}, the confronted models are consistent to each other for $0<\mid \triangle \text{AIC}\mid <2$. If $2\leq \mid \triangle \text{AIC}\mid <4$, there is certain disagreement between the confronted models. For $\mid \triangle \text{AIC}\mid \geq 4$, the model having large AIC value is disfavored by data. \\
For $0<\mid \triangle \text{BIC}\mid <2$, the model having large BIC value is weakly disfavored by the data. For $2\leq \mid \triangle \text{BIC}\mid < 6$, the model having large BIC value is strongly disfavored by the data. For $\mid \triangle \text{BIC}\mid\geq 6$, the model having large BIC value is very strongly disfavored by the data.\\ The summary of AIC and BIC values have been given in Table \ref{table:2}. For the CC data, $N=31$ and since $\Lambda$CDM (Generalized $\Lambda$CDM) model are having $H_0,\Omega_m$ ($H_0,m,n$) parameters respectively, the $\triangle \text{AIC}$ is highlighting that Generalized $\Lambda$CDM model has certain degree of disagreement with the $\Lambda$CDM model. For the joint analysis based on the Cosmic chronometer and Patheon data, one will have $N=1079$ and $0< \mid\triangle AIC\mid <2$, and thus these models may be said to be consistent with each other.\\ On the basis of $\mid \triangle\text{BIC}\mid$ value, we observe that BIC value of generalized $\Lambda$CDM model is greater than that of the $\Lambda$CDM model. In this sense, the generalized $\Lambda$CDM model is strongly (very strongly) disfavored by the Cosmic chronometer (joint) data respectively.
\begin{table}[htbp]
	\centering
	\begin{tabular}{|cc|c|c|c|c|c|}
		\hline
		Model &  dataset & $\chi^2_{min}$ & AIC & BIC & $\triangle \text{AIC}$ & $\triangle \text{BIC}$\\
		\hline
		$\Lambda$CDM & CC & $14.49363599$ & $18.9222$ &$21.3616$ & - & - \\
		 & CC+Pantheon & $1041.26977622$ & $1047.29$ &$1062.22$ & - & - \\
		\hline
		G-$\Lambda$CDM & CC & $14.48460821$ & $21.3735$ &$24.7866$ &$2.45129$ & $3.42496$ \\
		& CC+Pantheon & $1041.05674964$ & $1049.09$ &$1068.99$ & $1.80189$ & $6.77076$ \\
		\hline
	\end{tabular}
	\caption{The summary of model selection criteria}
	\label{table:2}
\end{table}

\section{Conclusions}\label{sec:7}
In this paper, we investigated the a cosmological model having homogeneous and isotropic line element with flat spatial sections and a parameterized form of Hubble parameter. This kind of Hubble parameter may interpolate between the decelerating past to the accelerating present of the universe. We show that that this kind of Hubble parameter may be a solution in the particle creation, bulk viscous, and $f(R)$ gravity framework of cosmological modeling for $\Gamma(H)=\alpha_1H+\frac{\alpha_2}{H}$, $\xi(H)=\beta_1+\beta_2H^2$ and $f(R)=a_1f(R)+a_2$ respectively where $\alpha_i,\beta_i,a_i,i=1,2$ are some constants containing constrained model parameters. The Raychaudhuri equation has been used to get the form of $f(R)$ function in the model.
\vspace{0.3cm}\\
We scrutinize the observational viability of considered Hubble parameter form to the Cosmic chronometer and Pantheon data. By using Bayesian statistical technique with MCMC analysis, we obtain model parameters's best fit values. The obtained best fit are $H_{0}=68.326^{+1.005}_{-1.045} \ Km/(s\cdot mpc)$, $m=0.307^{+0.059}_{-0.050}$, $n=3.040^{+0.165}_{-0.164}$ subjected to the CC data and $H_{0}=68.8^{+1.9}_{-1.9}  \ Km/(s\cdot mpc)$, $m=0.297^{+0.046}_{-0.076}$, $n=3.07^{+0.32}_{-0.32}$ subjected to the joint data of CC+Pantheon sample. At last, we find universe's present age is $t_{0}=13.52^{+1.73}_{-1.5}$ Gyr for CC data and $13.4^{+3.30}_{-2.04}$ Gyr for joint CC+Pantheon data. 
\vspace{0.3cm}\\
Furthermore, the behavior of cosmographic parameters suggest that the universe in model will behave like $\Lambda$CDM model in the limiting limits $z\rightarrow -1$. The early phase of the universe evolution is decelerating in nature which has been transitioned into the accelerating phase (see Figs. (\ref{fig:1}) and (\ref{fig:2})). According to model parameters best fit values, the transition red-shift is $z_{t}=0.63$. Additionally for our model, the present values of deceleration parameter are $q_{0}=-0.533$ (for the CC data) and $q_{0}=-0.5441$ (for the CC+Pantheon data). The universe is dominated by the cold dark matter-like component at large red-shifts and subsequently the universe expands under the influence of quintessence kind of dark energy and will eventually approaches the cosmological constant limit having  $\omega_{eff}=-1$ as $z \rightarrow -1$. 
\vspace{0.3cm}\\
It is evident in the model that the energy density will be decreasing  with time while preserving positive nature during the complete cosmological history. From the trends obtained according to the observational data, the value of jerk parameter decreases from early to late times and, finally approaches to $1$ which demonstrates that, in the early universe this model differs from the $\Lambda$CDM model and becomes similar to $\Lambda$CDM model in later times. Additionally, the jerk parameter's current values are $j=1.0186$ for the CC data and $j=1.0319$ for the CC+Pantheon data. In the early cosmos, the snap parameter ($s$) develops in the negative region. The  values of snap parameter at the present times are $s=-0.4467$ for the CC data and $j=-0.4482$ for the CC+Pantheon data. These calculated values of cosmographic parameters are consistent with the findings in the literature\cite{rev2}.
\vspace{0.3cm}\\
In summary, we show that the parameterized Hubble parameter cosmology may be a observationally viable one and, it may be admitted as a solution in the particle creation, bulk viscous, and $f(R)$ gravity framework also. The present generalized $\Lambda$CDM  model may not be favored over the $\Lambda$ cold dark matter model according to the Bayesian information criteria.

\section*{\textbf{Acknowledgments}}
G.P. Singh and A. Singh are thankful to the Inter-University Centre for Astronomy and Astrophysics (IUCAA), Pune, India for support under the Visiting Associateship programme. Authors are thankful to the honorable reviewer for highlighting different issues with suggestions.

\section*{\textbf{Data Availability Statement}}
This paper has no new associated data. All concepts as well as logical implications are stated in the paper with citations to the data sources.
\section*{\textbf{Declaration of competing interest}}
The authors declare that they have no known competing financial interests or personal relationships
that could have appeared to influence the work reported in this paper

\end{document}